\documentclass[reprint, amsmath, aps, amssymb, superscriptaddress]{revtex4-1}

\usepackage{hyperref}
\usepackage{braket}
\usepackage{graphicx}
\usepackage{dsfont}
\usepackage{mathtools}
\usepackage{units}

\begin{document}
\title{Isotropy and control of dissipative quantum dynamics}
\author{Benjamin Dive}
\affiliation{Department of Physics, Imperial College, SW7 2AZ London, UK}
\author{Daniel Burgarth}
\affiliation{Institute of Mathematics, Physics and Computer Science, Aberystwyth University, SY23 3FL Aberystwyth, UK}
\author{Florian Mintert}
\affiliation{Department of Physics, Imperial College, SW7 2AZ London, UK}
\date{\today}
\begin{abstract}
We investigate the problem of what evolutions an open quantum system described by a time-local Master equation can undergo with universal coherent controls. A series of conditions are given which exclude channels from being reachable by any unitary controls, assuming that the coupling to the environment is not being modified. These conditions primarily arise by defining decay rates for the generator of the dynamics of the open system, and then showing that controlling the system can only make these rates more isotropic. This forms a series of constraints on the shape and non-unitality of allowed evolutions, as well as an expression for the time required to reach a given goal. We give numerical examples of the usefulness of these criteria, and explore some similarities they have with quantum thermodynamics.

\end{abstract}
\maketitle

\section{Introduction}

The ability to coherently control quantum dynamics has received considerable interest in the last few decades, both for its potential application in technology \cite{Dowling2003, Dong2009, Schulte2012, Glaser2015} and the insight it provides to fundamental science \cite{Rabitz2000, Lambert2012, Goold2015}. It is therefore surprising that the question of what dynamics can be reached with unitary controls is poorly understood in open systems where interactions with the environment cannot be neglected \cite{Rivas2011, Breuer2002}. Current tools rely primarily on finding explicit solutions to control problems \cite{Machnes2011} but, as these methods are typically computational expensive and do not always give definitive answers, it is often hard to decide if the failure to find a good solution is due to its non-existence or simply an insufficient search. Having clear, efficiently accessible criteria which rules out certain evolutions avoids these problems and would help in the quest for improving the design and optimisation of devices for quantum computation, communication and sensing.

The question of what dynamics can be reached with coherent controls in the case of finite dimensional noiseless systems has been answered with the use of algebraic tools from the theory of Lie groups \cite{DAlesandro2008, Elliott2009a}. Attempts to generalise these methods to open systems have met considerable mathematical difficulties. The two principle results are an accessibility criterion \cite{Jurdjevic1972, Kurniawan2007, Kurniawan2012}, which describes which directions can be explored for short times; and Lie wedges \cite{Dirr2009, O'Meara2011}, which provide a partial characterisation of the geometry of the reachable set but in general cannot be calculated exactly. Other approaches focus on finding approximate numerical solutions \cite{Khaneja2005, Machnes2011}, or explore the related question of state-controllability, where the interest is in the ability to map one state to another \cite{Altafini2003, Altafini2004, Wu2006, Kurniawan2009, Bergholm2012, Sauer2014}. Yet another approach is to treat the system and environment on an equal footing and approach the infinite-dimensional problem directly \cite{Beauchard2010, Boscain2015}.

\begin{figure}[h]
\centering
\includegraphics[trim={7cm 14.9cm 12cm 10cm},clip, width=0.8\columnwidth]{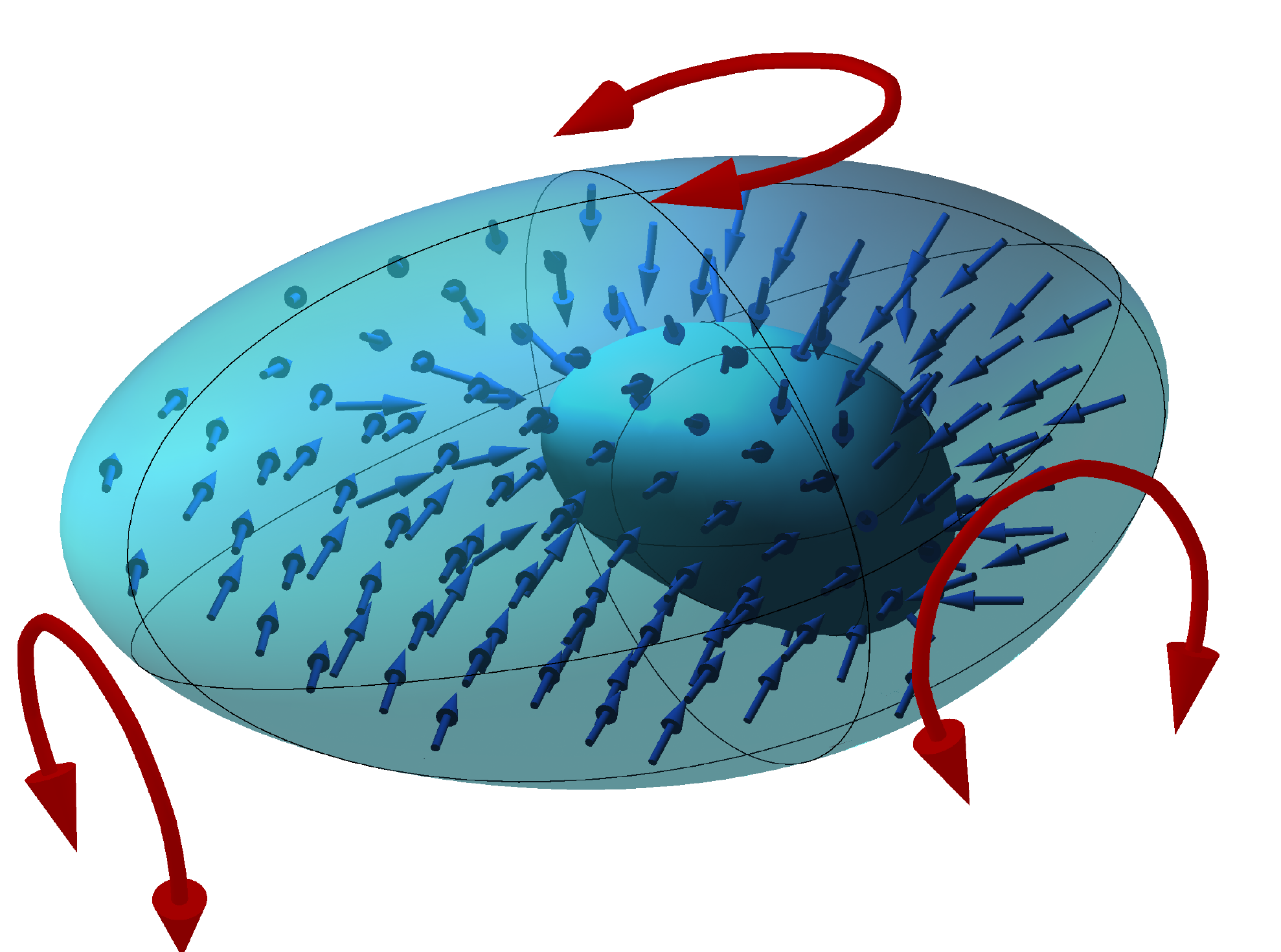}
\caption{We illustrate the principle ideas of this paper by showing a cross section of state space at two different times, with the small arrows indicating the direction of flow induced by a Lindbladian (the generator of memoryless dissipative dynamics). The large arrows correspond to the available controls. We see that different parts of the space are contracting at different rates; by rotating the system in time with Hamiltonian controls some of these decay rates can be averaged together.}
\label{fig:HighDimStatePlots}
\end{figure}

In this paper we investigate operator controllability from a geometric approach and characterise broad ranges of evolutions that a dissipative system cannot reach with \emph{any} unitary controls. In the case that there are experimental constraints that limit which of these controls could be physically realised, there are additional constraints imposed. The validity of the conditions detailed here is not affected however, as reducing the set of allowed operations can not increase what can be achieved. Furthermore our approach does not need a detailed knowledge of the behaviour of the environment or of the controls applied, but does require that the action of the environment on the system remains unchanged. Ideas similar to those we are presenting were introduced in \cite{Yuan2011} in the case of time-independent and unital quantum systems with a special focus on single qubits.

In order to get an intuitive understanding for the principle idea behind our work, it is useful to consider dissipative quantum processes as a flow in state space, as illustrated in Fig \ref{fig:HighDimStatePlots}.  The noise can act in a variety of ways on the state space, including rotating and shrinking the space (corresponding to decay) in a potentially anisotropic way. Hamiltonian controls allow us to impose additional rotations on the system such that different parts of the state space feel different contraction rates at different times. This results in the ability to mix the decay rates together and leads to the overall evolution obeying some averaged rates. These cause the final state space, which represents the total evolution, to be more isotropic than in the absence of controls.


Our main results relate directly to this, and state that a quantum operation cannot be reached if it has a more ordered structure than the noise acting on the system. After introducing the specific problem we are addressing and the relevant mathematics in section \ref{sec:setup}, we define the decay rates of an open system with this structure in mind in section \ref{sec:theoreticalresults}, and show that the action of any coherent controls is to make these more uniform. We demonstrate that the sum of these rates is unaffected by control and thus obtain a strict condition for the times at which a target evolution can be reached. In addition, we obtain bounds on how the purifying power of noise can be enhanced by control. The strength of these criteria is tested numerically in section \ref{sec:numresults} for common examples of noise, and we show that, at least for small systems, the necessary conditions are strong and tight enough to provide a major restriction on what evolutions are possible in realistic situations. Both the language and the mathematics used to describe these relations are reminiscent of thermodynamics, a link which we explore in section \ref{sec:thermo}. We conclude in \ref{sec:conclusion} with a summary of the results, a comparison of the methods of this paper with prior results on Lie wedges, and possible directions for future work.
 
\section{Problem and Mathematical Background}
\label{sec:setup}

The aim of this paper is to obtain some general rules for which operations cannot be performed on coherently controlled dissipative quantum systems by formalising the intuition described in Fig \ref{fig:HighDimStatePlots} and applying it to a more general setting. To do this we first state the control problem formally. We consider Hamiltonian controls on a finite dimensional system interacting with an environment, with the requirement that the reduced system obeys a time-local master equation
\begin{align}
\label{eq:EoM}
\frac{d}{dt}\rho &= G_t\left(\rho\right) = G_t^0(\rho) - i [H_t, \rho] \nonumber \\
\rho_T &= M_T(\rho_0)\equiv \mathcal{T}e^{\int_0^T dt G_t(\cdot)}\rho_0,
\end{align}
where $\rho$ is a quantum state, $G_t$ is the linear generator for the motion, $\mathcal{T}$ is the time-ordering operator, and $M_T$ is the resulting dynamical map, the set of which (varying over total times and controls) we aim to characterise. The generator is divided into an uncontrollable drift $G_t^0$ and a controllable Hamiltonian term $H_t$. The latter is a time-dependent control Hamiltonian chosen so as to generate the desired dynamics, and we impose no restrictions on it beyond being Hermitian. $G^0_t$ represents the intrinsic part of the dynamics, such as an internal energy splitting or an interaction with the environment. If we restrict it to be a Lindblad operator, which we will denote by $L_t$, then the allowed solutions are Markovian, completely-positive trace-preserving maps. Although the Lindbladian case is the most commonly used and the one with the clearest physical interpretation, the key results of this paper do not rely on the specific form of the Lindblad operator and hold for a more general generator which gives rise to non-Markovian dynamics. In such cases there can be substantial additional restrictions beyond the ones presented here, since the dynamics induced by control Hamiltonian and non-Markovian drift does not necessarily induce completely-positive dynamics, even if the uncontrolled evolution is completely-positive \cite{Lindblad1976}. The validity of the conditions presented in this paper are, however, not impaired by the additional intricacies of non-Markovian dynamics provided that the dynamics are still describable by a time-local linear generator.

An implicit assumption in Eq.(\ref{eq:EoM}) is that the control Hamiltonian does not affect the dissipative component of the generator. While there are cases where this approximation holds exactly \cite{Dalessandro2014}, this is not always the case, and it is possible for controls to modify the decoherence induced by the environment \cite{Arenz2014, Dive2015}. In such circumstances a variety of different methods have been developed \cite{Fischer2007, Rebentrost2009, Clausen2010, Clausen2011, Arenz2013}, but these typically require detailed knowledge about the environment or additional assumptions about finite-dimensionality or bounded interactions. They may also call for experimentally difficult regimes such as strong and/or rapidly oscillating control fields. For these reasons, it is highly desirable to explore what can be achieved with controls if the dissipative component of the dynamics is not modified, which is the regime studied in this paper. 

In order to make the picture introduced in Fig \ref{fig:HighDimStatePlots} rigorous, and to derive our results, it is necessary to introduce some mathematical concepts and notation. The starting point is to work in the generalised Bloch representation  \cite{Bengtsson2006}, where a quantum state $\rho$ is represented as the real vector $\ket{\rho} = (x_0, x_1, x_2,...,x_{d^2-1})^T$, where $x_i = \text{Tr}[\sigma_i \rho]$ are the expectation values over an orthonormal set of traceless Hermitian matrices for $i = 1,...,d^2-1$, $\sigma_0 = \tfrac{1}{\sqrt{d}}\mathds{1}$, and $d$ is the dimension of the underlying Hilbert space. In this representation super-operators acting on states become matrices. We will denote dynamical maps and generators in this representation by $\underline{M}$ and $\underline{G}$ respectively to distinguish them from their super-operator form. The spectral properties (eigenvalues, singular values, trace and determinant) of the super-operators are given by those of their matrix representation. The dual of a super-operator $M^\dagger$, defined according to $\text{Tr}[\mu M(\rho)] = \text{Tr}[M^\dagger(\mu) \rho]$, has as its matrix representation the Hermitian conjugate of the matrix representation of the original super-operator, such that $\underline{(M^\dagger)} = (\underline{M})^\dagger$.

Writing out the explicit form of $\underline{M}$ highlights some of its properties. When the dynamical map is trace-preserving $\underline{M}$ is of the form
\begin{equation}
\left(
\begin{array}{cccc}
 1 & 0 & 0 & ... \\
 v_1 & \tilde{\underline{M}}_{11} & \tilde{\underline{M}}_{12} & ... \\
 v_2 & \tilde{\underline{M}}_{21} & \tilde{\underline{M}}_{22} & ... \\
 ... & ... & ... & ... \\
\end{array}
\right)
\label{eq:matrixrep}
\end{equation}
where the top row is fixed, all the elements are real if $M$ is Hermiticity preserving (which it is for the vast majority of physically sensible cases), and the tilde refers to the reduced matrix. This form has the advantage of explicitly separating the unital and non-unital part of the dynamics. Unitality refers to leaving the maximally mixed state unchanged; as this is the only state left invariant by all Hamiltonians and it is the centre of rotations, this is an important property for control. The left hand column consisting of the elements $v_i$ fully describes the non-unital part of the map and quantifies how much the maximally mixed state is translated by the dynamical map. For unital operations, such as unitary evolution, these vanish and the dynamical map reduces to $\underline{M} = \underline{1}_1\oplus\tilde{\underline{M}}$. The reduced matrix $\tilde{\underline{M}}$ thus describes solely the unital part of the evolution. These are partially decoupled from the non-unital dynamics in the sense that $\widetilde{\underline{M}_B \underline{M}_A} = \tilde{\underline{M}}_B\tilde{\underline{M}}_A$, meaning that the total unital dynamics of a concatenation is given by the concatenation of the unital part of the individual super-operators. This can easily be seen by noting that the concatenation of super-operators is given by the product of their matrix representations.

As unitality is a key property, it is also useful to have a measure of how unital or non-unital a dynamical map is. A convenient one is $\text{Tr}\left[M(\tfrac{1}{d}\mathds{1})^2\right]$, the purity of the state obtained by applying the map to the maximally mixed state. This value is maximised at one if the maximally mixed state is mapped to a pure state and is minimised to $\tfrac{1}{d}$ if the map is unital. In a similar fashion the non-unitality can also be quantified by higher moments \cite{Sauer2014}, $\text{Tr}\left[M(\tfrac{1}{d}\mathds{1})^n\right]$, and the first $d$ moments are linearly independent. These should all be seen as describing roughly the same physical quantity: the ability of the map to purify states.

The matrix form of a trace-preserving generator $G$ is identical to Eq.(\ref{eq:matrixrep}), except that the entire top row vanishes. Its unital and non-unital parts can be separated in a similar way and the same results on concatenation holds. Because of this, we have that $\widetilde{e^{\underline{G}}} = e^{\tilde{\underline{G}}}$ and, hence, that if $M$ is the dynamical map generated by $G_t$ then $\tilde{M}$ is the one given by $\tilde{G}_t$. This means that contained inside every non-unital problem is a unital one with a dimension $1$ smaller, and any solution to the control problem of generating $M$ must also solve $\tilde{M}$. It is possible for this reduced problem to be non-Markovian even if the original one is Markovian, but this does not affect the validity of the approach.

As they are particularly important, we note the form that the super-operators of closed dynamics take in this representation. Unitary propagators, $M(\cdot) = U\cdot U^\dagger$, become matrices in the defining representation of the rotation group $\mathds{1}_1\oplus SO(d^2-1)$. Hamiltonian generators, $G(\cdot) = -i [H,\cdot]$, are in the corresponding Lie algebra, $0_1\oplus \mathfrak{so}(d^2-1$), which consist of real antisymmetric (and therefore traceless) matrices. In the case of $d=2$, unitary propagators form all such matrices (and like-wise for Hamiltonians), but in higher dimension they only form a subgroup. This means that in $d=2$ only, the set of every unitary on the system corresponds to all possible rotations of the state vector $\ket{\rho}$. In higher dimensions however, there exist rotations of this vector which cannot be induced by any Hamiltonian controls on the system; this is equivalent to saying that not every vector in the generalised Bloch space is a valid quantum state \cite{Bengtsson2006}. This property is one of the fundamental reasons why the results of this paper are necessary conditions rather than a complete characterisation of the allowed dynamics; we can only rule out targets which cannot be reached by any rotations, whether these are physical controls or not.

The intuitive picture described in the introduction relies on a notion of averaging a set (the decay rates) to obtain another. A natural way to describe this process is the majorization relation \cite{Bhatia1997} which tests if one real set is more uniformly distributed than an other. A set $\pmb{a}$ of real numbers is majorized by another such set $\pmb{b}$, written as $\pmb{a} \prec \pmb{b}$, if and only if
\begin{align}
a_1^\downarrow &\le b_1^\downarrow \nonumber\\
 a_1^\downarrow + a_2^\downarrow &\le b_1^\downarrow + b_2^\downarrow \\
&\;... \nonumber \\
\sum a_i^\downarrow &= \sum b_i^\downarrow, \nonumber
\end{align}
where $\downarrow$ signifies that the elements of the set are sorted in decreasing order. Another way of stating this is that $\pmb{a}$ is majorized by $\pmb{b}$ if and only if an ordering of $\pmb{a}$ can be obtained by a convex sum of different orderings of $\pmb{b}$. It is this property which makes it suitable to describe an averaging procedure. Another useful property is that it is conserved under scaling such that $\pmb{a} \prec \pmb{b}$ also implies $x \,\pmb{a} \prec x\,\pmb{b}$ for all real (including negative) $x$. A point to note is that, unlike standard inequalities on the reals, majorization provides only a partial order. \\

\section{Theoretical Results}
\label{sec:theoreticalresults}

With this formalism we are now in a position to refine the intuition developed in Fig \ref{fig:HighDimStatePlots}. The picture was of a Lindbladian (or a more general drift) acting on the state space such that it flowed from one shape to another. By coherently controlling the system, the space can be rotated so that some of the decay rates are averaged together. For a Markovian two-level system where the Lindbladian has no Hamiltonian component, these decay rates are the eigenvalues of the Lindbladian, which are always non-positive. In higher dimensions however these eigenvalues can be complex which gives rise to two problems as they may not faithfully quantify the contraction of the space and it is not clear what averaging them would signify. The situation is exacerbated in both the non-unital and non-Markovian case where there are even fewer constraints on the spectrum of the generator. This points to the need for a different way of quantifying the decay rates of an open system than naively taking the spectrum of the drift.

Working through the mathematics in detail (as we do in section \ref{sec:proofs}) shows that the correct decay rates to consider are the eigenvalues of the sum of the drift and its dual, or equivalently, of the Hermitian part of $\underline{G}$. Their physical relevance is supported by two important properties. Firstly, as the generators of rotations in $\underline{G}$ are anti-Hermitian (whether it corresponds to a Hamiltonian degree of freedom or not), these rates capture only the decay/growth component of the flow. Secondly they are real, so averaging them corresponds to a more uniform flow in a way which can be naturally defined using majorization. These suggest that the eigenvalues of the Hermitian part of the drift capture some of the key aspects of the controllability of the system, and their prominence in the criteria detailed below show that this is indeed the case. In a similar fashion it can be seen that the anisotropy of the dynamical map is described by its singular values (as they are rotationally invariant and non-negative), which loosely correspond to the characteristic lengths of the final state space.

This, together with majorization as described above, allows us to write the conditions which must all be satisfied for a dynamical map $M$ to be reachable by a system with drift $G_t^0$. These are: an expression for the state space volume reached at a given time (independent of the controls),
\begin{equation}
\label{eq:detTimeCondition}
\det{\left(M\right)} = e^{\int_0^T \text{Tr}\left(G_t^0\right) dt},
\end{equation}
which extends prior work on the determinant of quantum channels \cite{Wolf2008a}; a constraint on the anisotropy of the dynamical map,
\begin{equation}
\label{eq:majorizationrelation}
\log{[\pmb{\sigma}(M)]} \prec \int_0^T \pmb{\lambda}\left(\frac{G_t^0 + G_t^{0\dagger}}{2}\right) dt,
\end{equation}
where $\pmb{\lambda}$ and $\pmb{\sigma}$ refer to the set of eigenvalues and singular values of the super-operator respectively, and the $\log$ acts element-wise on the set; a unital version of this condition (for trace-preserving drifts),
\begin{equation}
\label{eq:unitalmajorizationrelation}
\log{[\pmb{\sigma}(\tilde{M})]} \prec \int_0^T \pmb{\lambda}\left(\frac{\tilde{G}_t^0 + \tilde{G}_t^{0\dagger}}{2}\right) dt,
\end{equation}
which was shown in \cite{Yuan2011} for a time-independent Lindbladian; and bounds on the maximal non-unitality that can be reached,
\begin{align}
\label{eq:nonunitalitycondition}
\text{Tr}[M(\tfrac{1}{d}\mathds{1})^n] \le \sup_{\rho} \left\{ \text{Tr}[\rho^n] \;|\; \exists\;t : \text{Tr}[\rho^{n-1} G_t^0(\rho)] = 0\right\}
\end{align}
which gives rise to independent conditions for $n=2,...,d$. In order to highlight the systematic similarities of the conditions, and stress that it is the Hermitian part of the drift that matters (which in the Bloch representation removes any Hamiltonian or other rotational component), $G_t^0$  can be replaced by $\tfrac{G_t^0 + G_t^{0\dagger}}{2}$ in Eqs.(\ref{eq:detTimeCondition}) and (\ref{eq:nonunitalitycondition}).

Instead of seeing if the system can reach a target map, we can instead ask if a time-independent generator $G^0$ and controls can approximately simulate another generator $G^\prime$ arbitrarily well. In this case we obtain the condition
\begin{equation}
\label{eq:generatormajorization}
\pmb{\lambda}\left(G^\prime + G^{\prime\dagger}\right) \prec \pmb{\lambda}\left(G^0 + G^{0\dagger}\right)
\end{equation}
whose unital version
\begin{equation}
\pmb{\lambda}\left(\tilde{G}^\prime + \tilde{G}^{\prime\dagger}\right) \prec \pmb{\lambda}\left(\tilde{G}^0 + \tilde{G}^{0\dagger}\right)
\label{eq:unitalgeneratormajorization}
\end{equation}
also holds provided both generators are trace-preserving.

We proceed to give a proof of these conditions, followed by a detailed discussion.

\subsection{Proofs}
\label{sec:proofs}

\paragraph*{Evolution time ---} To derive Eq.(\ref{eq:detTimeCondition}), we begin by noting that the formal solution for $\underline{M}$ is given in terms of time-ordered matrix exponential, which can be expressed according to the Magnus expansion \cite{Blanes2009}
\begin{align}
\label{eq:Magnus}
\underline{M} &= \mathcal{T}e^{\int_0^T \underline{G}(t) dt} \nonumber \\
&= e^{\int_0^T \underline{G}(t_1) dt_1 + \tfrac{1}{2} \int_0^t dt_1\int_0^{t_1} dt_2 [\underline{G}(t_1),\underline{G}(t_2)] + ... }
\end{align}
where all higher order terms in the series consist of nested commutators. As the determinant of a matrix exponential is the exponential of the trace we can rewrite this as
\begin{align}
\det{(\underline{M})} &=  e^{\text{Tr}\left[\int_0^T \underline{G}(t_1) dt_1 + \tfrac{1}{2} \int_0^t dt_1\int_0^{t_1} dt_2 [\underline{G}(t_1),\underline{G}(t_2)] + ...\right] } \nonumber \\
&= e^{\text{Tr}\left[\int_0^T \underline{G}(t) dt\right]}
\end{align}
where we have used the fact that commutators are traceless. As discussed previously, the control Hamiltonian appear in the equation of motion \eqref{eq:EoM} as commutators, therefore their Bloch representation are also traceless giving $\text{Tr}[\underline{G}(t)] = \text{Tr}[\underline{G}^0(t)]$. As the trace and determinant of $M$ and $G$ are identical to those of $\underline{M}$ and $\underline{G}$, this gives the desired expression
\begin{equation}
\det{(M)} = e^{\int_0^T \text{Tr}\left[G^0_t\right] dt}.
\end{equation}
In the case that the Magnus expansion does not converge (which may happen if $\int_0^T || \underline{G}(t) ||_2 dt > \pi$  \cite{Blanes2009}), then the 
proof can be extended by splitting the propagator into sufficiently many terms
\begin{equation}
\underline{M} = \mathcal{T}e^{\int_{t_n}^T \underline{G}(t) dt} ... \mathcal{T}e^{\int_{t_1}^{t_2} \underline{G}(t) dt} \mathcal{T}e^{\int_0^{t_1} \underline{G}(t) dt}
\end{equation}
such that the Magnus expansion converges for each term. Applying the same steps as before to each term and using the fact that the determinant of a product is the product of the determinants we arrive at
\begin{align}
\det{(M)} &= \det{(\mathcal{T}e^{\int_{t_n}^T \underline{G}(t) dt})}...\det{(\mathcal{T}e^{\int_0^{t_1} \underline{G}(t)})} \nonumber \\
&= e^{\int_{t_n}^T \text{Tr}\left[G^0_t\right] dt}...e^{\int_{t_1}^{t_2} \text{Tr}\left[G^0_t\right] dt}e^{\int_0^{t_1} \text{Tr}\left[G^0_t\right] dt} \nonumber \\
&= e^{\left\{\int_{t_n}^T \text{Tr}\left[G^0_t\right] dt + ... + \int_{t_1}^{t_2} \text{Tr}\left[G^0_t\right] + \int_{t_0}^{t_1} \text{Tr}\left[G^0_t\right] dt\right\}} \\
&= e^{\int_0^T \text{Tr}\left[G^0_t\right] dt} \nonumber
\end{align}
as before.\\

\paragraph*{Anisotropy of the dynamical map ---} The proofs of Eqs.(\ref{eq:majorizationrelation}-\ref{eq:unitalmajorizationrelation}) arise from two observations. Firstly, the evolution can always be decomposed into infinitesimal time-steps in a Trotter-like way, alternating between coherent and incoherent evolution. Secondly, the controls only affect the coherent steps which are all rotation matrices and so do not modify the singular values of the incoherent time steps, as singular values of a matrix depend only on the product of that matrix with its Hermitian adjoint. To prove Eq.(\ref{eq:majorizationrelation}) we expand the time-ordered exponential in terms of short time steps
\begin{align}
\underline{M} &= \mathcal{T} e^{\int_0^T \underline{G}(t) dt}, \nonumber \\
\label{eq:TrotterLike}
&= \lim_{\delta t\to0} \left( e^{\underline{G}^0(T)\delta t}e^{\underline{H}(T)\delta t} ... \,e^{\underline{G}^0(0)\delta t}e^{\underline{H}(0)\delta t}\right).
\end{align}
We now consider the singular values of both sides of the equation, denoted by the operator $\pmb{\sigma}$. Specifically, we use \cite{Bhatia1997}
\begin{align}
\log\pmb{\sigma}(AB) \prec \log\pmb{\sigma}(A) + \log\pmb{\sigma}(B)
\end{align}
for the majorization relation between the singular values of matrices and their products, where the log is understood as acting on each element in the set, and the sum on the right hand side acts on the elements of the sets ordered by magnitude. Generalising this to the case of multiple sums and applying it to Eq.(\ref{eq:TrotterLike}), we obtain
\begin{align}
\log\pmb{\sigma}(\underline{M}) \prec \lim_{\delta t\to0} \Big\{ &\log\pmb{\sigma}\left(e^{\underline{G}^0(T)\delta t}e^{\underline{H}(T)\delta t}\right) + ... \\
& ... + \log\pmb{\sigma}\left(e^{\underline{G}^0(0)\delta t}e^{\underline{H}(0)\delta t}\right)\Big\}. \nonumber
\end{align}
As mentioned above, the coherent steps corresponds to rotation matrices and therefore do not affect the singular values. This allows the expression for the singular values to be simplified to
\begin{align}
\label{eq:majorizationexpansion}
\log\pmb{\sigma}(\underline{M}) \prec \lim_{\delta t\to0} \left\{ \log[\pmb{\sigma}(e^{\underline{G}^0(T)\delta t})] +...+ \log[\pmb{\sigma}(e^{\underline{G}^0(0)\delta t})] \right\}.
\end{align}
We recall that singular values are obtained by $\pmb{\sigma}(A) = \pmb{\lambda}\left(\sqrt{AA^\dagger}\right)$, where $\pmb{\lambda}$ signifies the eigenvalues. From this, each term in the previous equation can be expressed for small $\delta t$ as
\begin{align}
\label{eq:majorizationeigenvalueexp}
\log\pmb{\sigma}(e^{\underline{G}^0(t)\delta t}) &= 
\log\left[ \pmb{\lambda}\left( e^{\underline{G}^0(t)\delta t} e^{\underline{G}^{0\dagger}(t) \delta t} \right)^{\tfrac{1}{2}}\right], \nonumber \\
& \approx \log\left[ \pmb{\lambda}\left( e^{(\underline{G}^0(t) + \underline{G}^{0\dagger}(t))\delta t +[\underline{G}^0(t), \underline{G}^{0\dagger}(t)] \delta t^2}\right)^{\tfrac{1}{2}}\right], \nonumber \\
&= \tfrac{1}{2}\pmb{\lambda}\left((\underline{G}^0(t)+ \underline{G}^{0\dagger}(t)) \delta t + [\underline{G}^0(t), \underline{G}^{0\dagger}(t)] \delta t^2 \right)
\end{align}
where higher order terms can be calculated using the Baker-Campbell-Hausdorff formula. The first term is of order $\delta t$ and, as the number of terms in Eq.(\ref{eq:majorizationexpansion}) is $\tfrac{T}{\delta t}$, it contributes to the integral in the limit $\delta t \to 0$ while all the higher order terms vanish. This gives as the final expression
\begin{align}
\log{[\pmb{\sigma}(\underline{M})]} &\prec \lim_{\delta t \to 0} \Big\{ \pmb{\lambda}\left(\frac{\underline{G}^0(T) + \underline{G}^{0\dagger}(T)}{2}\right)\delta t + ... \nonumber \\
&\;\;\;\;\;\;\;\; ... + \pmb{\lambda}\left(\frac{\underline{G}^0(0) + \underline{G}^{0\dagger}(0)}{2}\right)\delta t + \frac{T}{\delta t}O(\delta t^2) \Big\}, \nonumber \\
&\prec \int_0^T \pmb{\lambda}\left(\frac{\underline{G}^0(t) + \underline{G}^{0\dagger}(t)}{2}\right) dt,
\end{align}
which is independent of the representation used and so holds for the super-operators themselves. The proof for condition (\ref{eq:unitalmajorizationrelation}), the unital version of this for trace-preserving generators, follows immediately from the fact that $\underline{M} = \mathcal{T} \exp\{\int_0^T \underline{G}(t) dt\}$ implies $\underline{\tilde{M}} = \mathcal{T} \exp\{\int_0^T \underline{\tilde{G}}(t) dt\}$, as was discussed in section \ref{sec:setup}. A different proof of this latter unital result was shown in \cite{Yuan2011} for time-independent Lindbladians only, and relied on similar mathematical ideas.\\

\paragraph*{Maximal non-unitality ---} To prove the non-unitality bounds, Eq.(\ref{eq:nonunitalitycondition}), we begin with the formal expression and then derive an easily evaluable bound for it. To do this we are required to make the additional assumption that $G_t$ is continuous. The maximal non-unitality of an open system is quantified by
\begin{align}
\label{eq:maxnonunital}
\text{Tr}[M(\tfrac{1}{d}\mathds{1})^n] &\le \sup_{t, H_\tau} \text{Tr}[\rho^n(t, H_\tau)] \\
&\text{ where } \;\rho(t,H_\tau) = \mathcal{T} e^{\int_0^t (G_\tau^0+H_\tau) (\cdot) d\tau}\tfrac{1}{d}\mathds{1}, \nonumber
\end{align} 
and the supremum is over all possible evolution times and all possible controls. It is sufficient to consider only $H_\tau$ which are defined for $\tau \in [0,\infty)$. Eq.(\ref{eq:maxnonunital}) appears as difficult to calculate as solving the control problem, and is therefore of limited use. However an upper bound for it can be found more readily. To do this we note a property that the supremum must satisfy as a function of $t$ for any $H_\tau$. From this, we reformulate the constraint that $\rho$ has been evolved from the maximally mixed state into one which is easier to work with.

For a given continuous $H_\tau$ finding the supremum of Eq.(\ref{eq:maxnonunital}) reduces to finding the supremum of a scalar function which is bounded between $\tfrac{1}{d}$ and $1$ and differentiable everywhere. There are several cases in which this could happen. Firstly, the supremum being reached at $t=0$ can be immediately excluded as the function is a minimum at that point. Secondly, the supremum being reached for some finite time leads to $\tfrac{d}{dt}\text{Tr}[\rho^n(t)] = 0$ at that point in time. Lastly, if the supremum is not reached for finite $t$ then either: it is reached in the limit $t\to\infty$ and so the derivative also goes to $0$ in this limit (due to the function being bounded from above), or the limit is undefined because the function does not converge. In the latter case there are many local maxima which form a series, the supremum of which gives the supremum of the original function. As the gradient of each these maxima is $0$, the largest value that can be reached by the function also occurs when the gradient vanishes. Hence, we have that a necessary condition for the supremum of Eq.(\ref{eq:maxnonunital}) is
\begin{align}
\tfrac{d}{dt}\text{Tr}&[\rho^n] = 0,\nonumber \\
\text{Tr}&[\rho^{n-1}\,G^0_t (\rho)] -i\text{Tr}\left[\rho^{n-1}\,[H_t, \rho]\right] = 0, \nonumber \\
\text{Tr}&[\rho^{n-1}\,G^0_t (\rho)] -i\text{Tr}\left[\,[H_t, \rho^n]\,\right] = 0, \nonumber \\
\text{Tr}&[\rho^{n-1}\,G^0_t (\rho)] = 0.
\end{align}
Since the gradient as calculated above depends solely on $G^0_t$ and not on the controls, we can relax the condition on $G_t$ being continuous to $G^0_t$ being continuous. Instead of calculating the supremum over all controls, we can compute it over all states which satisfy this condition. This allows us to place a bound on the maximum non-unitality reachable by a dynamic system which, as desired, does not require any propagators to be calculated 
\begin{align}
\text{Tr}[M(\tfrac{1}{d}\mathds{1})^n]  \le \sup_{\rho} \left\{ \text{Tr}[\rho^n] \;|\; \exists\;t : \text{Tr}[\rho^{n-1} G_t^0(\rho)] = 0\right\}.
\end{align}
This provides up to $d$ constraints (including the trivial case for $n = 1$), as higher moments of $\rho$ are not independent. It is interesting to note that similar expressions were also arrived at in a different control problem, that of finding the possible steady states of a driven open system \cite{Sauer2014}.\\

\paragraph*{Generator anisotropy ---} The criterion for generator anisotropy (applicable only to time-independent generators) is similar to the one for the anisotropy of the dynamical map, with the important difference that the target is a flow in state space which we desire to achieve continuously in time, rather than a snapshot of the evolution at a single instance. The derivation for this condition begins with Eq.(\ref{eq:majorizationrelation}), where we replace the target $M$ by $e^{G^\prime T}$, and limit ourselves to the system drift being time-independent, such that our starting point is
\begin{equation}
\log{[\pmb{\sigma}(e^{G^\prime T})]} \prec \int_0^T \pmb{\lambda}\left(\frac{G^0 + G^{0\dagger}}{2}\right) dt \;\;\forall \;T.
\label{eq:startgenani}
\end{equation}
If this condition is satisfied for infinitesimal $\delta t$ then, by concatenation, it holds for all time $T$ and $G^0$ can effectively simulate $G^\prime$. We use the term `effectively' to emphasise that although the generator $G^\prime$ cannot be reached exactly, it is possibly to follow a trajectory in state space which is arbitrarily close to the one generated by it. Under these conditions Eq.(\ref{eq:startgenani}) simplifies to
\begin{align}
\log{\pmb{\sigma}(e^{G^\prime \delta t})} &\prec \tfrac{1}{2} \pmb{\lambda}\left(G^0 + G^{0\dagger}\right) \delta t, \nonumber \\
\log{\pmb{\lambda}(e^{G^\prime\delta t}e^{G^{\prime\dagger}\delta t})^\frac{1}{2}} &\prec \nonumber \\
\log{\pmb{\lambda}\left(e^{(G^\prime+G^{\prime\dagger}) \frac{\delta t}{2}+ O(\delta t^2) }\right)} &\prec \nonumber \\
\pmb{\lambda}\left(\frac{G^\prime + G^{\prime\dagger}}{2}\delta t +O(\delta t^2) \right) &\prec \\
\implies \pmb{\lambda}\left(G^\prime + G^{\prime\dagger}\right) &\prec \pmb{\lambda}\left(G^0 + G^{0\dagger}\right). \nonumber
\end{align}
The unital version of this relation also holds provided both generators are trace-preserving for the reasons discussed in section \ref{sec:setup}. We note that this condition implies that $\text{Tr}[G^\prime] = \text{Tr}[G^0]$ is also required.

There are cases where the time it takes to simulate the dynamics is not of concern, which corresponds to the traces of $G$ and $G^\prime$ not being equal. In such cases, the condition above can be relaxed to
\begin{align}
\tfrac{1}{2 \text{Tr}[G^\prime]}\pmb{\lambda}\left(G^\prime + G^{\prime\dagger}\right) &\prec \tfrac{1}{2 \text{Tr}[G^0]}\pmb{\lambda}\left(G^0 + G^{0\dagger}\right)
\end{align}
by a rescaling of time in Eq.(\ref{eq:startgenani}). This comes by replacing $T$ with $\tfrac{\text{Tr}[G]}{\text{Tr}[G^\prime]}T$ on the left hand side that that equation. This is a relaxation of Eq.(\ref{eq:majorizationrelation}), as it holds even if $\text{Tr}[G^\prime] \ne \text{Tr}[G^0]$.\\

\paragraph*{Unital qubit Lindbladians ---} For the case of qubits undergoing unital Lindbladian dynamics, Eq.(\ref{eq:generatormajorization}) can be simplified to \cite{Yuan2011}
\begin{equation}
\label{eq:eq:unitalqubitlind}
\pmb{\lambda}\left(L^\prime\right) \prec \pmb{\lambda}\left(L^0\right)
\end{equation}
as the dissipative part of the Bloch representation of any unital qubit Lindbladian is symmetric \cite{Lendi1987} and drift Hamiltonians are not of interest to us (they can be cancelled out by controls). This criterion is also sufficient. To prove it, we provide an explicit way to reach $\tilde{L}^\prime$ using a drift $\tilde{L}_0$ and unrestricted Hamiltonian controls. For simplicity, we pick the time scale of the target dynamics such that $\text{Tr}[\tilde{L}^\prime] = \text{Tr}[\tilde{L}_0] = 1$. By using the singular value decomposition, the target can be expressed as
\begin{align}
\underline{\tilde{M}} = e^{\underline{\tilde{L}}^\prime t} = U D V = U
\left(\begin{matrix}
  e^{-\nu_1 t} & 0 & 0\\
  0 & e^{-\nu_2 t} & 0\\
  0 & 0 & e^{-\nu_3 t}
\end{matrix}\right)
V
\end{align}
where the $-\nu_i$ are the eigenvalues of $\underline{\tilde{L}}^\prime$, and $U$ and $V$ are elements of $O(3)$. Furthermore, as $\underline{M}$ has a positive determinant and the diagonal block is positive, we can pick $U$ and $V$ to have determinant $+1$, thereby restricting them to $SO(3)$. In a similar way, we can express the free evolution of the system for time $t$ as
\begin{align}
e^{\underline{\tilde{L}}_0 t} = W F(t) W^\dagger = W
\left(\begin{matrix}
  e^{-\mu_1 t} & 0 & 0\\
  0 & e^{-\mu_2 t} & 0\\
  0 & 0 & e^{-\mu_3 t}
\end{matrix}\right)
W^\dagger
\end{align}
where the $-\mu_i$ are the eigenvalues of $\underline{\tilde{L}}_0$. Using the same argument as above, $W$ can be chosen to be in $SO(3)$. Controls on the system allow the implementation of any $R = e^{\underline{\tilde{H}}}$ which, as we noted previously, corresponds to any matrix in $SO(3)$. The control scheme to reach the target map corresponds to alternating free evolution and instantaneous controls as
\begin{equation}
\label{eq:SolveUnitalQubit}
U D V = R_1 W F(t_1) W^\dagger R_2 .... W F(t_n) W^\dagger R_{n+1}
\end{equation}
We pick $R_1 = U W^\dagger$, $R_{n+1} = W V$ and relabel $W^\dagger R_j W = R'_j$ which can always be done due to the group structure. Next we pick the $R'$ to be permutation matrices (which all lie in SO(3)) such that Eq.(\ref{eq:SolveUnitalQubit}) consists solely of diagonal matrices where every term is an exponential. This lets us express the previous matrix equation in the simple form
\begin{align}
\label{eq:convexcombinationsingulars}
\left(\begin{matrix}
\nu_1 \\ \nu_2 \\ \nu_3
\end{matrix}\right) = 
\left(\begin{matrix}
\mu_1 \\ \mu_2 \\ \mu_3
\end{matrix}\right) t_1 + 
\left(\begin{matrix}
\mu_1 \\ \mu_3 \\ \mu_2
\end{matrix}\right) t_2 + ... + 
\left(\begin{matrix}
\mu_3 \\ \mu_2 \\ \mu_1
\end{matrix}\right) t_6,
\end{align}
where, we recall from the way we picked the scale of $L_0$, that $\sum t_i = 1$. This control scheme thus allows us to reach any $\underline{\tilde{M}}=e^{\underline{\tilde{L}}^\prime t}$ where the eigenvalues of $\underline{\tilde{L}}^\prime$ are a convex combination of those of $\underline{\tilde{L}}^0$; which is equivalent to saying that they are majorized by them \cite{Bhatia1997}. This means that $\pmb{\lambda}\left(L^\prime\right) \prec \pmb{\lambda}\left(L^0\right)$ is a sufficient, as well as necessary, condition for reachability in unital qubit systems with unconstrained Hamiltonian control.\\

\subsection{Discussion}
\label{sec:discussion}

\paragraph*{Evolution time ---} The first criterion, Eq.(\ref{eq:detTimeCondition}), is an equality which appeared in \cite{Wolf2008a} for the Markovian and time-independent case. Extending it to the time-dependent case gives it an important use in control theory: it states that a target map may only be reached by a dynamical system for the times which satisfy Eq.(\ref{eq:detTimeCondition}). To understand why this is the case we note that the modulus of the determinant of $M$ is the volume occupied by its image and the trace of $G_t$ is the rate at which this state space is growing (this will be non-positive, unless the system is non-Markovian, leading to a contraction of the space). The interpretation of this result is thus that the total rate at which volume is lost in state space is independent of the Hamiltonian controls. More insight can be gained by noting that the trace of the drift is always positive in the physically sensible case of the generator being Hermiticity preserving. This means that the evolution of such a system can only reach maps with positive determinant \cite{Wolf2008a, Dirr2009}. As this is not the case for every completely-positive trace-preserving map, this condition allows us to immediately rule out large sections of the space as unreachable for a broad class of dynamical systems.

If we further restrict the drift to be a Lindbladian (with a non-vanishing dissipative part) at all times, then the trace is always negative, signifying that a target map can only be reached at a single instant in time (if at all), and that this time can be easily calculated as it is independent of the controls. If the drift is non-Markovian (where the interplay between memory effects and controls has received much recent attention \cite{Machnes2014, Reich2014}) the trace of the drift can be positive for certain times, leading to revivals in the determinant. Indeed, this has already been suggested as an indicator of non-Markovianity \cite{Lorenzo2013}. For our purposes, this particular feature leads to the possibility of there being several solutions to Eq.(\ref{eq:detTimeCondition}) for a given target map. In both cases, the precise information about required evolution time given by this condition is in stark contrast to the case of closed systems, where in general very little is known about the time required to reach a target without explicitly solving for the evolution of the system. \\

\paragraph*{Anisotropy of the dynamical map ---} The conditions of Eqs.(\ref{eq:majorizationrelation}) and (\ref{eq:unitalmajorizationrelation}) are a refinement of the intuition that controlling the system allows us to average the decay rates of the drift together. This is most easily seen by considering the unital case. The left hand side of the relation are the singular values of the dynamical map which, in the same way as the determinant is the volume in state space, are the characteristic lengths of the final state space. Thus, while Eq.(\ref{eq:detTimeCondition}) determines the volume reached, Eq.(\ref{eq:unitalmajorizationrelation}) provides a constraint on the anisotropy of the dynamical maps that can be reached.

The non-unital majorization relation has broadly the same interpretation, although the overall shift caused by the non-unitality manifests itself in the decay rates and singular values in a complex way. Indeed, one of the eigenvalues of $G_t^0 + G_t^{0\dagger}$ will typically be positive in the non-unital case, even if the drift is Markovian. While the idea of a positive decay rate in a Markovian system may appear counterintuitive, it only signifies that some states become purer under such a Lindbladian. This is most easily seen by considering the Bloch sphere: negative eigenvalues correspond to states moving towards the maximally mixed state, but if the system is decaying to the state $\ket{0}$ then there is also a dynamic evolution away from the centre towards a pure state on the surface of the sphere. Thus, despite their similar form, Eqs.(\ref{eq:majorizationrelation}) and (\ref{eq:unitalmajorizationrelation}) give very different results and there are many dynamical maps that satisfy one but not the other for a given drift (as we will show in Fig \ref{fig:LGAD}). It is also worth noting that Eq.(\ref{eq:detTimeCondition}) is recovered, up to a modulus, by the last term in the majorization relations.\\

\paragraph*{Maximal non-unitality ---} This last condition on dynamical maps, Eq.(\ref{eq:nonunitalitycondition}), is conceptually very different from the others. Rather than restricting the shape of the dynamical map, it provides a series of constraints on how much the maximally mixed state can be displaced, corresponding to where the centre of the image of the dynamical map lies in the state space. Although the right hand side of Eq.(\ref{eq:nonunitalitycondition}) is independent of controls, the maximisation over all states (and over all $t$ if the drift is time-dependent) makes this criterion somewhat harder to evaluate in higher dimensions. It is worth noting that the constraint is less strict than $G_t^0(\rho) = 0$, which means that the non-unitality is not bounded by the fixed points of the drift. The interpretation of this criterion is therefore that it is possible to increase the ability of noise to purify states by using controls, but only up to the limits given.

A physical example of this is a three-level system in a $\Lambda$ configuration (such as in Fig \ref{fig:LambdaSys}), with the excited state decaying into the two low level states. In the absence of controls, the system has some non-unitality as the maximally mixed state over the three levels will decay to a mixed state over only two levels. With the use of controls, however, the population can be coherently transferred back from one of the two ground states to the excited state where it will once again decay. Doing this many times results in the total population being transferred to the other ground state and the total action of the dynamics is to map everything to a single pure state. Thus, this dynamical map induced by a specific set of controls has maximal non-unitality. This is the principle behind optical pumping and shows that non-unitality can be increased with controls. If the two lower levels had some decay between them, however, this scheme may not work perfectly and Eq.(\ref{eq:nonunitalitycondition}) provides bounds for how well it can be done.\\

\paragraph*{Generator anisotropy ---} Instead of investigating if the system can reach a target map, in Eq.(\ref{eq:generatormajorization}) we consider if it can be made to approximate a different drift continuously in time. To do this we limit ourselves to time-independent $G_t^0$ and see if it can give rise to evolutions arbitrarily close to $M = e^{G^\prime t}$ for all $t$. If it can, we say that $G^0$ can effectively simulate $G^\prime$ as it can replicate the same dynamics arbitrarily well in a time continuous fashion. Necessary but not sufficient criteria to do this are given by Eqs.(\ref{eq:generatormajorization}) and (\ref{eq:unitalgeneratormajorization}). These are stricter than the anisotropy conditions on dynamical maps; it imposes not only a target map but the whole trajectory in time to it. That this can be done at all is at first hand surprising, as the only generators which can be reached exactly from a given drift are given precisely by the drift plus all possible controls. However, by quickly rotating the system it is possible to get arbitrarily close to the required trajectory by winding tightly around the desired path without ever moving exactly along it.

In the case of the generator $G^0$ being a unital qubit Lindbladian, Eq.(\ref{eq:generatormajorization}) can be simplified further to $\pmb{\lambda}\left(L^\prime\right) \prec \pmb{\lambda}\left(L^0\right)$ and is \emph{sufficient} \cite{Yuan2011}. Furthermore the eigenvalues of such unital qubit Lindbladians, $\pmb{\lambda}(L)$, are constrained by complete positivity \cite{Gorini1976}. From this it is straightforward to show that $\pmb{\lambda}(L) = -(\tfrac{1}{2}, \tfrac{1}{2}, 0)$ majorizes the spectrum of all other such Lindbladians. Hence, one with such a spectrum, such as dephasing $L(\cdot) = -[\sigma_z, [\sigma_z,\,\cdot\,]\,]$, is universal and it can simulate all other unital qubit Lindbladians. Conversely, the completely depolarising channel with eigenvalues $\pmb{\lambda}(L) = -(\tfrac{1}{3}, \tfrac{1}{3}, \tfrac{1}{3})$ is majorized by all other Lindbladians and is therefore at the bottom of the hierarchy defined by majorization.\\

\section{Numerical Results}
\label{sec:numresults}

The conditions detailed above are almost all necessary but not sufficient, so the question of how tight they are is important. We carried out numerical simulations to quantify this in two ways: firstly by how often the criteria forbid a target from being reached, and secondly by how often a target can be reached when it is not excluded. The issue with doing this is that, for the very reason that the criteria derived in this paper are useful, it is computationally very difficult to test if a dynamical map can be reached with a given drift and controls. The only definite method requires simulation and optimisation of the control problem. The size of the simulation itself scales as $d^4$, and the cost of optimising the control pulses scales far worse \cite{Machnes2011}. Nevertheless, we obtained results for a class of common non-unital qubit and qutrit Lindbladians.

\begin{figure}[t]
\includegraphics[scale=0.8]{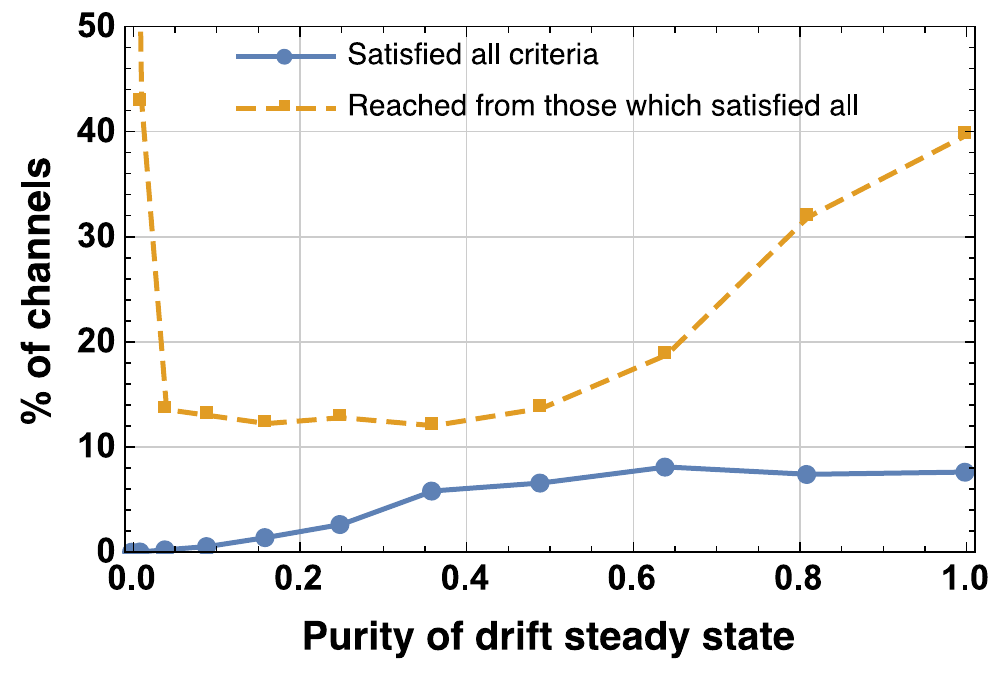}
\includegraphics[scale=0.8]{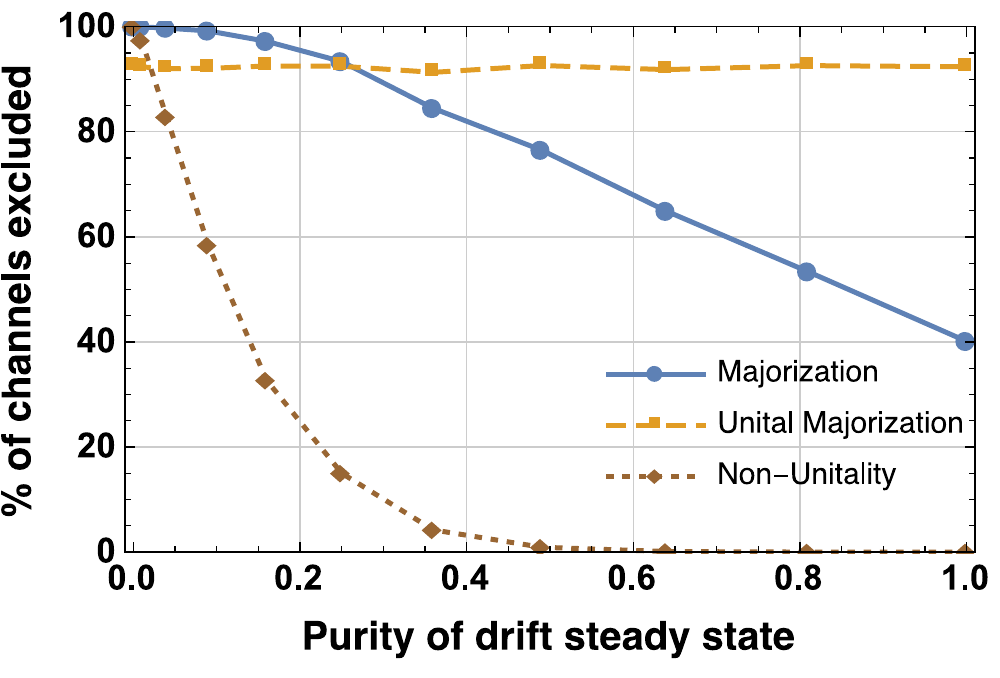}	
\caption{The graphs characterise the strength of Eqs.(\ref{eq:majorizationrelation}-\ref{eq:nonunitalitycondition})  for generalised amplitude damping Lindbladians as a function of the purity of the drift steady state - itself a function of the bath temperature. The top graph shows what fraction of the randomly generated channels satisfied all of Eqs.(\ref{eq:majorizationrelation}-\ref{eq:nonunitalitycondition}) and, out of those, how many could be reached with a numerical optimisation package \cite{Qutip} to within a distinguishability (given by the diamond norm) of at least $0.1\%$. The bottom graph shows what fraction of the same channels are ruled out by each of the conditions individually. }
\label{fig:LGAD}
\end{figure}

For the qubit system the Lindbladian we consider is generalised amplitude damping, a ubiquitous type of noise, corresponding to a qubit which can exchange an excitation with a bath at finite temperature \cite{Nielsen2000}. This can be thought of as a spin which has a finite rate for transitioning from the excited to the ground state and from the ground to the excited state, where the ratio between the two is a function of temperature. The temperature determines the non-unitality of the noise: at zero temperature the steady state is the pure ground state, while at infinite temperature it is the maximally mixed state. As the non-unitality of the Lindbladian is an important aspect of its controllability, we use the purity of the steady state as a parametrisation of temperature. We analysed the ratio of randomly generated time-dependent Markovian maps \cite{Wolf2008} which could be reached numerically, and whether they satisfied the criteria Eqs.(\ref{eq:majorizationrelation})-(\ref{eq:nonunitalitycondition}), for different values of the non-unitality of the drift. Eq.(\ref{eq:detTimeCondition}) was used in deciding the evolution time for which we attempted to find solutions of the control problem.

The results for such a Lindbladian at different temperatures are shown in Fig \ref{fig:LGAD}. Taken together, the criteria state that over $90\%$ of the space is unreachable at each temperature considered, showing that the conditions are useful as they cut out the large majority of dynamical maps as impossible to achieve. The insufficiency of the criteria manifests itself in that - at some temperatures - only $10\%$ of those not ruled out can be reached. This number, however, approaches $100\%$ in the unital case, which is expected as we know that the majorization condition is sufficient in unital qubit systems. That this figure rises again for highly non-unital, low temperature baths shows that the criteria are increasingly useful in this limit too. It is also interesting to note that the relative importance of the different conditions varies with temperature: when the noise has a pure fixed point the unitality criterion provides no information and the unital majorization criterion is the most restrictive, while their importance is reversed when the fixed point is maximally mixed.

\begin{figure}[b]
\includegraphics[trim = {3cm 22cm 3cm 1cm}, clip, width=0.75\columnwidth]{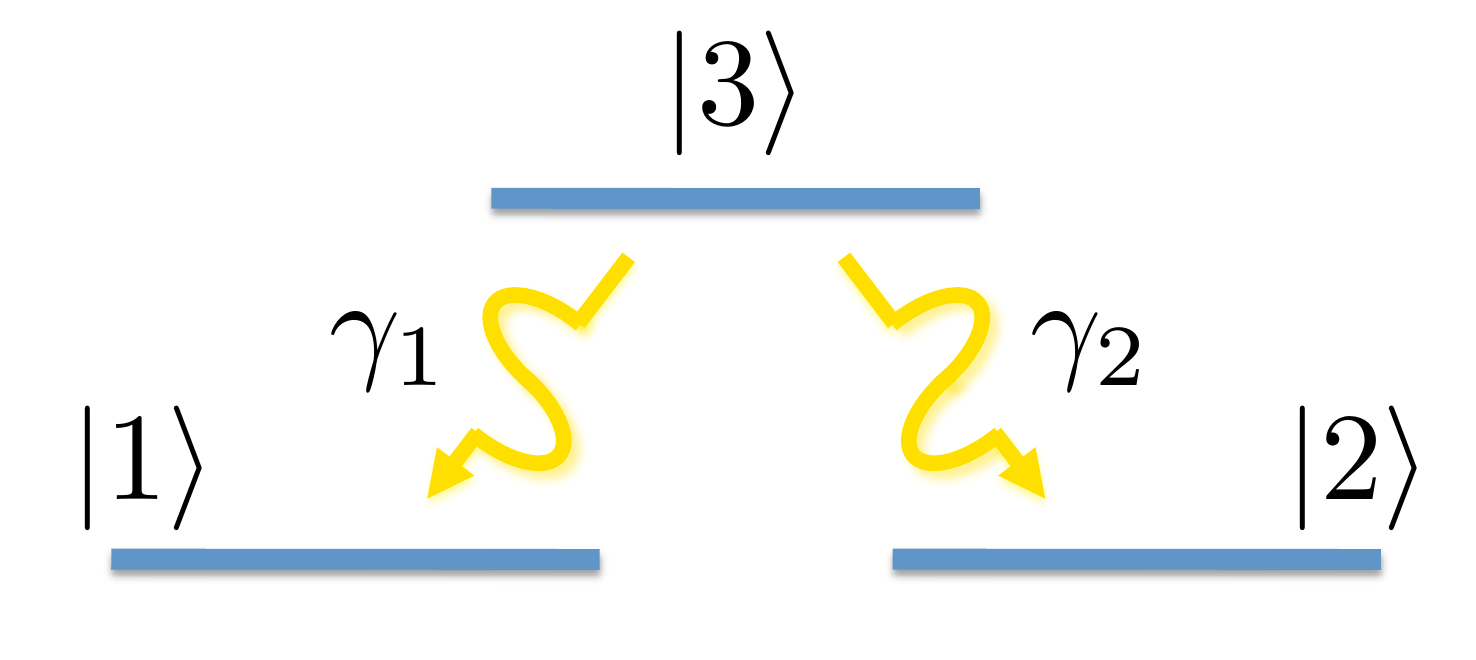}
\caption{Sketch of a qutrit in a $\Lambda$ configuration, where the Lindbladian causes the excited level to decay to the two ground state at different rates. The skew, $\gamma_1/\gamma_2$, characterises the asymmetry in the system.}
\label{fig:LambdaSys}
\end{figure}

\begin{figure}[t]
\includegraphics[scale=0.8]{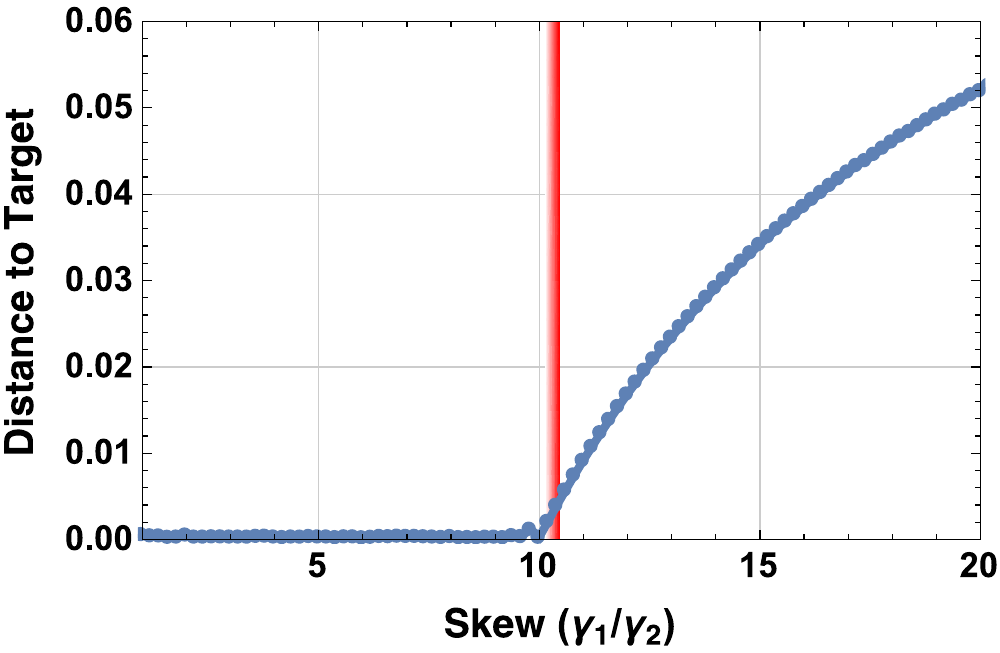}
\caption{
The graph shows how the dynamical map anisotropy conditions relates to the controllability of the non-symmetric $\Lambda$ system outlined in figure \ref{fig:LambdaSys}. The system drift had a skew of $10$ and we attempted to reach maps generated by the same Lindbladian but with skews between $1$ and $20$. Plotted is the minimal distance (given by the diamond norm) to targets with different skews that could be reached with a numerical optimisation package \cite{Qutip}. The thick red line at $10.3$ is the non-reachable boundary given by the majorization criteria - everything to the right of it is excluded - in excellent agreement with the numerical results. The slight bump just below 10 is due to the difficulty of numerically finding solutions which require rapidly oscillating control Hamiltonians.}
\label{fig:LambdaPlotLin}
\end{figure}

The second example is another common type of noise, sketched in Fig \ref{fig:LambdaSys}, is a qutrit in a non-symmetric $\Lambda$ configuration where the top level decays to the two lower levels according to the Lindbladian
\begin{equation}
L(\rho) = \sum_{i=1,2} \gamma_i \left( L_i \rho L_i^\dagger - \tfrac{1}{2} \{L_i^\dagger L_i\,,\,\rho\}\right)
\end{equation}
where $L_i = \ket{i}\bra{3}$ \cite{Schirmer2004}. We focus on how the controls and the skew, $\gamma_1/\gamma_2$, can influence the asymmetry of the final evolution. To do this we picked a drift with a fixed skew, and investigated how close the system could get to maps generated by a similar drift but with a different skew. Fig \ref{fig:LambdaPlotLin} shows that those which were as or more symmetric (a skew closer to $1$) as the drift could be reached with a very high fidelity, and increasingly poorly those which were less symmetric. This is in excellent agreement with the majorization criteria as plotted. The tightness of the necessary conditions in this scenario show how useful they are in cases where there is a clear measure of non-uniformity, demonstrating that we can use controls to go from a highly ordered evolution to a less ordered one, but not the other way around.

This result may at first hand appear to contradict the conclusion arrived at in section \ref{sec:discussion} when this example was also discussed on maximal non-unitality in the context of optical pumping. There we said that a qutrit in a $\Lambda$ configuration could have a pure fixed point regardless of the ratio of the decay rates; while here we stress that the skew cannot be increased. The resolution of this problem is that although the fixed point of the dynamics can be chosen independently of the skew, this only determines the evolution at $t\to\infty$, at all other times the state space occupies a finite volume and the shape of this volume is what is constrained by the skew.

We expect that in general the tightness of the criteria would decrease as the system size increases. The main argument for this is that the number of rotations in the state space that do not correspond to Hamiltonian degrees of freedom grows with dimension. That the conditions were tighter in the qutrit than the qubit example, however, suggests that there are specific cases where they remain an excellent approximation to the allowed operations. Even when less tight, the potential usefulness of the conditions may be greater for larger systems as it is substantially harder to simulate these and therefore to learn about their controllability via other means.\\

\section{Relations to Thermodynamics}
\label{sec:thermo}

The focus of this paper has been the majorization conditions (\ref{eq:majorizationrelation}-\ref{eq:unitalmajorizationrelation}) and (\ref{eq:generatormajorization}-\ref{eq:unitalgeneratormajorization}) which provide limits on the operations that can be reached on an open quantum system. It is interesting to note that majorization also plays a key role on the related question of state, rather than operator, controllability. A central result there is that a state $\sigma$ can be reached from a state $\rho$ by a unital completely-positive trace-preserving map if and only if $\pmb{\lambda}(\sigma) \prec \pmb{\lambda}(\rho)$ \cite{Bengtsson2006}. This is a stricter form of the Second Law of Thermodynamics as the majorization relation $\pmb{\lambda}(\sigma) \prec \pmb{\lambda}(\rho)$ imposes the constraint $S(\sigma) \ge S(\rho)$ on the von-Neumman entropy \cite{Nielsen2000} of the two states, but the converse is not always true. This means that, under unital evolution, the eigenvalues of a state cannot become more ordered and therefore the entropy cannot decrease.

The natural extension of this to the results on generator controllability, Eqs.(\ref{eq:generatormajorization}) and (\ref{eq:unitalgeneratormajorization}), is that it lifts a form of the Second Law from applying to states to super-operators. This is a restatement of what we have shown: that generators of dissipative dynamics cannot become more ordered by the presence of coherent controls. These relations imply that the process is irreversible, controls can be used to make an existing generator of noise $G^0$ arbitrarily close to a different generator $G^\prime$, but the reverse cannot be done even approximately. This is a surprising result as the controls themselves are fully reversible as they are coherent. The rise of irreversibility from purely reversible pieces is a long standing puzzle of quantum mechanics and a key aspect of the Second Law, the criteria we have developed here shows that it applies to super-operators as well as states. 

A more explicit link between the present results and thermodynamics can be found by considering the rate of change of the entropy of a system as it undergoes unital evolution. An expression for this in terms of the spectral properties of the channel is given in \cite{Raginsky2002} which can be easily modified for Markovian channels, using Eq.(\ref{eq:majorizationrelation}), to give
\begin{equation}
\label{eq:entropyproduction}
\frac{d}{dt}S[\rho(t)] \ge \frac{\lambda_1}{2} ||\rho(t)-\tfrac{\mathds{1}}{d}||_2^2
\end{equation}
where $\rho(t)$ is a state evolving under a unital Lindbladian (possibly under the presence of controls), $||\cdot||_2$ is the $L_2$ norm, and $\lambda_1$ is the smallest (in magnitude) eigenvalue of the unital part of the Hermitian part of the Lindbladian. While the left hand side is any channel generated by a Lindbladian and control, the lower bound is the smallest decay rate of the Lindbladian and independent of the controls. This shows that the minimal rate of entropy production cannot be lowered. The physical picture is that rotating a system as it decays cannot increase how well states are shielded from the production of entropy.\\

\section{Conclusion}
\label{sec:conclusion}

The principle idea behind this work is that the decay rates of the generator of an open system, the eigenvalues of the sum of the generator and its dual, provides limitations as to the operations the system can achieve with coherent controls, resulting in Eqs.(\ref{eq:detTimeCondition}-\ref{eq:unitalgeneratormajorization}). The decay rates can be made more isotropic by coherent controls, corresponding to the rates being averaged out by rotations, but the total rate of decay cannot be changed and it is not possible to create a more ordered structure or to increase the non-unitality beyond a given limit. These hold for a range of open quantum systems - going beyond Markovian ones - within some assumptions which are discussed earlier.

These assumptions are shared with existing work on the controllability of Lindbladians based on Lie wedges \cite{Dirr2009, O'Meara2011}. The Lie wedge provides a sufficient but not necessary condition for controllability (as the semigroup closure still needs to be taken), while the criteria of this paper are necessary but not sufficient. Taken together, they allow us to approximate the reachable set from both sides. Our approach gives results which are easier to use and can be calculated numerically, while in many cases there are no known methods to determine the exact Lie wedge, especially in the non-unital case. It also has the considerable advantage of allowing drifts which are non-Markovian and time-dependent. However, the method used here does not enable us to see what effect reducing the allowed set of controls has. In the simplest case of unital qubit Lindbladians, we saw that majorization was sufficient if we had unrestricted Hamiltonian controls; a result which can also be obtained from Lie wedges, showing the consistency of the two methods.

The results we have presented are a useful tool in the quest for designing quantum systems to achieve desired non-unitary tasks, as they rule out some dynamics as impossible without the high cost of simulation and optimisation. Two examples of the use of the criteria were investigated numerically highlighting that, although they are necessary but not sufficient, they still give a practical approximation to the allowed operations. Due to the partial order induced by majorization, it shows that some types of noise are ``more useful'' than others as they can be used to replicate all the same evolutions, in addition to others. Directions to develop this further include applying it to different types of systems, such as Gaussian ones. Another avenue would be to investigate the effect of reducing the allowed set of controls, for example by considering the system as multi-partite where only controls local to the sub-systems are possible. Further investigations into the link with thermodynamics may also prove fruitful.\\

\paragraph*{Acknowledgement --- } We would like to thank the QuTIP project \cite{Qutip} for providing the numerical packages used, and especially to Alex Pitchford for developing the pulse optimisation package for it. We are grateful to HPC Wales for providing the computational time required. Fruitful discussions were had with David Jennings, Kamil Korzekwa and Matteo Lostaglio on the links with thermodynamics, for which we are thankful. This work was supported by EPSRC through the Quantum Controlled Dynamics Centre for Doctoral Training, the EPSRC grant EP/M01634X/1, and the ERC project ODYCQUENT.\\

\bibliography{/Users/bd313/library}
\end{document}